\documentstyle[12pt,amssymb,amsmath]{article}
\newcommand{\p}{\partial}
\newcommand{\la}{\lambda}
\newcommand{\La}{\Lambda}

\begin{document}

\title{ {\bf Constructing $N$-soliton solution for the mKdV equation through constrained flows
 }  }
\author{ {\bf  Yunbo Zeng\dag \footnote {E-mail:yzeng@math.tsinghua.edu.cn}
 \hspace{1cm} \hspace{1cm} Huihui Dai\ddag\hspace{1cm}} \\
    {\small {\it \dag
 Department of Mathematical Sciences, Tsinghua University,
 Beijing 100084, China}} \\
    {\small {\it \ddag
 Department of Mathematics, City University of Hong Kong, 
 Kowloon, Hong Kong, China}}  }  
\date{}
\maketitle
\renewcommand{\theequation}{\arabic{section}.\arabic{equation}}

\begin {abstract}
Based on the factorization of soliton equations into two commuting integrable 
$x$- and $t$-constrained flows, we derive $N$-soliton solutions for mKdV equation 
via its $x$- and $t$-constrained flows. It shows that soliton solution for 
soliton equations can be constructed directly from the constrained flows.
\end{abstract}

\hskip\parindent

 
\hskip\parindent

\section{Introduction}
\setcounter{equation}{0}
\hskip\parindent

It is well known that there are several methods to derive the $N$-soliton
solution of soliton equations, such as the inverse scattering method,
the Hirota method, the dressing method, the Darboux transformation, etc.
(see, for example, \cite{1,2,3} and references therein). In present paper, we
propose a method to construct  $N$-soliton solution for mKdV equation directly through 
two commuting $x$- and $t$-constrained
flows obtained from the factorization of mKdV equation. It was shown in \cite{4,5,6,7} that
(1+1)-dimensional soliton equation can be factorized by $x$- and
$t$-constrained flow which can be transformed into two commuting $x$- and
$t$-finite-dimensional integrable Hamiltonian systems. The Lax representation
for constrained flows can be deduced from the adjoint representation of the
auxiliary linear problem for soliton equations \cite{8}. By means of the Lax
representation and  the standard method in \cite{9,10,11} we are able to introduce the
separation variables for constrained flows \cite{12}-\cite{16} and to establish
their Jacobi inversion problem \cite{14,15,16}. Furthermore, the factorization of soliton
equations and separability of the constrained flows allow us to find the
Jacobi inversion problem for soliton equations \cite{14,15,16}. By using the Jacobi
inversion technique \cite{17,18}, the $N$-gap solutions in term of Riemann theta
functions for soliton equations can be obtained, namely, the constrained
flows can  be used to derive the $N$-gap solution. The present paper   shows that 
the $x$- and $t$-constrained
flows and their Lax representation can also  be used to directly construct  
the $N$-soliton solution for soliton equations. In fact the method proposed 
in this paper together with that in the previous paper \cite{19} provides 
a general procedure to derive $N$-soliton solution for soliton equations via 
their constrained flows.
\ \
  
\section{The factorization of the mKdV  hierarchy}
\setcounter{equation}{0}
\hskip\parindent

We first briefly recall  the constrained flows of the mKdV hierarchy 
and their Lax representation.
The mKdV hierarchy 
\begin{equation}
\label{s1}
 q_{t_{2n+1}}=D b_{2n+1}=D\frac{\delta H_{2n+1}}{\delta q},
 \qquad n=0,1,\cdots,
\end{equation}
with 
$$H_{2n+1}=\frac {2a_{2n+2}}{2n+1},$$
is associated with the reduced AKNS spectral problem for $r=-q$ \cite{1}
\begin{equation}
\label{s2}
 \left( \begin{array}{c} 
 \psi_1 \\ \psi_2 \end{array} \right)_x=
 U \left( \begin{array}{c}
 \psi_1 \\ \psi_2 \end{array} \right),
 \qquad
 U = \left( \begin{array}{cc}
 -\lambda & q \\ -q&\lambda\end{array} \right),
\end{equation}
and the evolution equation of the eigenfunction
\begin{equation}
\label{s3}
 \left( \begin{array}{c} 
 \psi_1 \\ \psi_2 \end{array} \right)_{t_{2n+1}}=
 V^{(2n+1)}(q, \lambda) \left( \begin{array}{c}
 \psi_1 \\ \psi_2 \end{array} \right),
\end{equation}
where
\begin{equation}
\label{s4}
 V^{(2n+1)}=\sum_{j=0}^{2n+1} \left( \begin{array}{cc}
 a_j & b_j \\ c_j & -a_j \end{array} \right)\lambda^{2n+1-j},   
\end{equation}
with
$$a_0=-1,\quad b_0=c_0=a_1=0, \quad b_1=-c_1=q,\quad 
a_2=-\frac 12q^2,\quad b_2=c_2=-\frac 12q_x,\quad ...,$$
and in general
\begin{eqnarray}
\label{s5}
& b_{2m+1}=-c_{2m+1}=Lb_{2m-1}, \quad
L=\frac 14D^2+qD^{-1}q D,\quad D=\frac {d}{dx},\quad DD^{-1}=D^{-1}D=1, \nonumber \\ 
& b_{2m}=c_{2m}=-\frac 12Db_{2m-1}, \qquad a_{2m+1}=0, \qquad
 a_{2m}=2D^{-1}qb_{2m}.
\end{eqnarray}

For the well-known mKdV equation
\begin{equation}
\label{s6}
 q_{t}=D b_{3}=\frac 14(q_{xxx}+6q^2q_x),
\end{equation}
the $V^{(3)}$  reads
\begin{equation}
\label{s7}
 V^{(3)}=\left( \begin{array}{cc}
 -\la^3-\frac 12 q^2\la &q\la^2-\frac 12q_x\la+\frac 14q_{xx}+\frac 12q^3 
\\-q\la^2-\frac 12q_x\la-\frac 14q_{xx}-\frac 12q^3  & \la^3+\frac 12 q^2\la 
\end{array} \right).   
\end{equation}

We have
\begin{equation}
\label{s8}
\frac{\delta \lambda}{\delta q}=\psi_1^2+\psi_2^2, \qquad
L(\psi_1^2+\psi_2^2)
=\lambda^2 (\psi_1^2+\psi_2^2).
 \end{equation}

The x-constrained flows of the mKdV hierarchy  consist
 of the equations obtained from
 the spectral problem  (\ref{s2}) for $N$
distinct real numbers $\lambda_j$ and the restriction of the variational
derivatives for the conserved
quantities $H_{2k_0+1}$ (for any fixed $k_0$) and $\lambda_{j}$defined by 
(see, for example, \cite{4}-\cite{7}, \cite{20,21})
\begin{subequations}
\label{s9}
\begin{equation}
\label{s9a}
\psi_{1j,x}=-\lambda_j\psi_{1j}+q\psi_{2j},\qquad
\psi_{2j,x}=-q\psi_{1j}+\lambda_j\psi_{2j},\qquad
 j=1,\cdots,N,
\end{equation}
\begin{equation}
\label{s9b}
\frac{\delta H_{2k_0+1}}{\delta q}-\frac 12
 \sum_{j=1}^N \frac{\delta \lambda_j}{\delta q}
\equiv b_{2k_0+1}-\frac 12 \sum_{j=1}^N(\psi_{1j}^2+\psi_{2j}^2)=0.
\end{equation}
\end{subequations}

For $k_0=0$, (\ref{s9b}) gives
\begin{equation}
\label{s10}
q=\frac 12(<\Psi_1, \Psi_1>+<\Psi_2, \Psi_2>),
\end{equation}
where
$$\Psi_k=(\psi_{k1},\cdots,\psi_{kN})^{T},\qquad k=1,2,
\qquad  \Lambda=diag
(\lambda_1,\cdots,\lambda_N).$$
 By substituting (\ref{s10}),  (\ref{s9a}) becomes a finite-dimensional 
integrable Hamiltonian system (FDIHS)
$$\Psi_{1x}=-\La \Psi_1+\frac 12(<\Psi_1, \Psi_1>+<\Psi_2, \Psi_2>)\Psi_2
=\frac {\p \overline {H}_0}{\p \Psi_2},$$
\begin{equation}
\label{s11}
\Psi_{2x}=-\frac 12(<\Psi_1, \Psi_1>+<\Psi_2, \Psi_2>)\Psi_1+\La \Psi_2
=-\frac {\p \overline {H}_0}{\p \Psi_1},
\end{equation}
with
$$\overline {H}_0=-<\La \Psi_1, \Psi_{2}>+\frac 18(<\Psi_1, \Psi_1>+<\Psi_2, \Psi_2>)^2.$$
Under the constraint (\ref{s10}), the t-constrained flow obtained from 
(\ref{s3}) with $V^{(3)}$ given by (\ref{s7}) for $N$ distinct $\la_j$ 
can also be written as a FDIHS
\begin{equation}
\label{s12}
\Psi_{1t}=\frac {\p \overline {H}_1}{\p \Psi_2},\qquad
\Psi_{2t}=-\frac {\p \overline {H}_1}{\p \Psi_1},
\end{equation}
with
$$\overline {H}_1=-<\La^3 \Psi_1, \Psi_{2}>
-\frac 18(<\Psi_1, \Psi_1>+<\Psi_2, \Psi_2>)^2<\La\Psi_1, \Psi_{2}>$$
$$+\frac 14(<\Psi_1, \Psi_1>+<\Psi_2, \Psi_2>)(<\La^2 \Psi_1, \Psi_1>
+<\La^2 \Psi_2, \Psi_{2}>)-\frac 18<\La\Psi_1, \Psi_1>^2$$
$$-\frac 18<\La\Psi_2, \Psi_2>^2+\frac 14<\La\Psi_1, \Psi_1><\La\Psi_2, \Psi_2>
+\frac 1{128}(<\Psi_1, \Psi_1>+<\Psi_2, \psi_2>)^4.$$

The Lax representation for the 
constrained flows (\ref{s11}) and (\ref{s12}), which can be obtained from 
the adjoint representation of the Lax representation for mKdV hierarchy \cite{6,8}, 
is  given by
  $$M_x=[\widetilde U, M],\qquad M_{t}=[\widetilde V^{(3)}, M]$$
where $\widetilde U$ and $\widetilde V^{(3)}$ are obtained from $U$ and $V^{(3)}$ 
by inserting (\ref{s10}) and the Lax matrix $M$ is of the form
$$ M=\left( \matrix A(\la)&B(\la)\\C(\la)&-A(\la)\endmatrix\right),\qquad
A(\la)=-\la-\sum_{j=1}^{N}\frac{\la\la_j\psi_{1j}\psi_{2j}}
{\la^2-\la^2_{j}},$$
$$B(\la)=\frac 12(<\Psi_1, \Psi_1>+<\Psi_2, \Psi_2>)
+\frac 12\sum_{j=1}^{N}\frac{\la_j}{\la^2-\la^2_{j}}
[(\la+\la_j)\psi^2_{1j}-(\la-\la_j)\psi^2_{2j}],$$
$$C(\la)=-\frac 12(<\Psi_1, \Psi_1>+<\Psi_2, \Psi_2>)
+\frac 12\sum_{j=1}^{N}\frac{\la_j}{\la^2-\la^2_{j}}
[(\la-\la_j)\psi^2_{1j}-(\la+\la_j)\psi^2_{2j}]. $$

The compatibility of (\ref{s2}), (\ref{s3}) and (\ref{s1}) ensures that  if $\Psi_1, \Psi_2$
satisfies two commuting FDIHSs (\ref{s11}) and (\ref{s12}), simultaneously, then $q$
given by (\ref{s10}) is a solution of mKdV equation (\ref{s6}), namely, the mKdV equation
(\ref{s6}) is factorized by the $x$-constrained flow (\ref{s11}) and $t$-constrained
flow (\ref{s12}).

\section{Constructing the $N$-soliton solution for the mKdV  equation}
\setcounter{equation}{0}
\hskip\parindent

 Hereafter we assume that  $q(x,t), \psi_{1j}, \psi_{2j}$ be real functions. 
For soliton solution we have $q(x,t)\rightarrow 0, \psi_{1j}\rightarrow 0, 
\psi_{2j}\rightarrow 0$, when $|x|\rightarrow \infty.$ In order to obtain convenient 
formulas to construct $N$-soliton solution, we need to rewrite all the formulas by 
using the complex version instead of the vector version. We denote
$$\Phi=\Psi_1+i\Psi_2,\qquad \phi_j=\psi_{1j}+i\psi_{2j}.$$
Then (\ref{s11}) and (\ref{s12}) become
\begin{equation}
\label{q1}
\Phi_x=-\Lambda\Phi^*-\frac i2\Phi^T\Phi^*\Phi,
\end{equation}
\begin{equation}
\label{q2}
\Phi_{t}=-\La^3\Phi^*-\frac i2\Phi^T\Phi^*\La^2\Phi+\frac i2\La\Phi^*\Phi^T\La\Phi
-\frac i2\Phi\Phi^T\La^2\Phi^*,
\end{equation}
where we have used $\overline {H}_0=0$.

The generating function of integrals of motion for the system (\ref{q1}) and
(\ref{q2}),  $\frac 12Tr M^2(\la)=A^2(\la)+B(\la)C(\la)$, gives rise to
$$A^2(\la)+B(\la)C(\la)=\la^2-2 \overline H_0+
\sum_{j=1}^{N}\frac{F_{j}}{\la^2-\la^2_{j}}, $$
where $F_j, j=1,...,N,$ are $N$ independent integrals of motion for the
systems (\ref{q1}) and
(\ref{q2})
$$F_j=2\la_j^3\psi_{1j}\psi_{2j}-\frac 12\Phi^T\Phi^*\la_j^2(\psi_{1j}^2+\psi_{2j}^2)
+\frac 14\la_j^2(\psi_{1j}^2+\psi_{2j}^2)^2
+\frac 12\sum_{k\neq j}\frac{\la_j^2}{\la^2_j-\la^2_{k}}P_{kj},$$
$$P_{kj}=\la_j\la_k(4\psi_{1j}\psi_{2j}\psi_{1k}\psi_{2k}+\psi_{1j}^2\psi_{1k}^2
+\psi_{2j}^2\psi_{2k}^2-\psi_{1j}^2\psi_{2k}^2-\psi_{2j}^2\psi_{1k}^2)$$
$$-\la^2_k(\psi_{1j}^2\psi_{1k}^2
+\psi_{2j}^2\psi_{2k}^2+\psi_{1j}^2\psi_{2k}^2+\psi_{2j}^2\psi_{1k}^2), \quad j=1,...,N.$$

Using (\ref{q1}), we have
$$P_{kj}=-\frac 12[\la_k\phi_{k}\phi^*_{j}(\la_k\phi^*_{k}\phi_{j}-\la_j\phi_{k}\phi^*_{j})+
\la_k\phi_{j}\phi^*_{k}(\la_k\phi^*_{j}\phi_{k}-\la_j\phi_{j}\phi^*_{k})],$$
$$\la_j\phi_{j}\phi^*_{j}\p_x^{-1}(\phi^2_{j}+{\phi^*}^2_{j})=-(\phi_{j}\phi^*_{j})^2,$$
\begin{equation}
\label{q3}
\la_k\phi_{j}\phi^*_{k}-\la_j\phi_{k}\phi^*_{j}=
(\la^2_j-\la^2_{k})\p_x^{-1}(\phi_{j}\phi_{k}).
\end{equation}
In a similar way as we did in \cite{19}, in order to constructing $N$-soliton solution, 
we have to set $F_j=0$. By using (\ref{q1}) and (\ref{q3}) $F_j$ can be rewritten as
$$F_j=\frac i2\la_j^2\phi_{j}^*[-\phi_{jx}
+\frac i2\sum_{k=1}^N\la_k\phi_k\p_x^{-1}(\phi_{j}\phi_{k})]-\frac i2\la_j^2\phi_{j}
[-\phi^*_{jx}-\frac i2\sum_{k=1}^N\la_k\phi^*_k\p_x^{-1}(\phi^*_{j}\phi^*_{k})]=0
, $$ 
which leads to
$$\phi_{jx}=-\gamma_j\phi_j+\frac i2\sum_{k=1}^N\la_k\phi_k\p_x^{-1}(\phi_{j}\phi_{k}),
\quad \quad j=1,...,N,$$
or equivalently
\begin{equation}
\label{q4}
\Phi_{x}=-\Gamma \Phi+\frac i2\p_x^{-1}(\Phi\Phi^T)\La\Phi=-\Gamma \Phi+R\Phi, 
\end{equation}
where $\Gamma=\text{diag}(\gamma_1,...,\gamma_N), \gamma_j$ are undetermined real numbers and
\begin{equation}
\label{q5}
R=\frac i2\p_x^{-1}(\Phi\Phi^T)\La.
\end{equation}
Notice that
\begin{equation}
\label{q6}
\frac i2\Phi\Phi^T=R_x\La^{-1},\qquad
\La R=R^T\La,\end{equation}
it follows from (\ref{q4}) and (\ref{q5}) that
$$R_x=\frac i2\p_x^{-1}(\Phi_x\Phi^T+\Phi\Phi_x^T)\La$$
\begin{equation}
\label{q7}
=\p_x^{-1}(-\Gamma R_x+RR_x-R_x\Gamma+R_xR)=-\Gamma R-R\Gamma+R^2.
\end{equation}
We now show that
$\Gamma=\La.$
In fact, it is found from (\ref{q4}) and (\ref{q7}) that
$$\Phi_{xx}=-\Gamma \Phi_x+R\Phi_x+R_x\Phi
=-\Gamma(-\Gamma\Phi+R\Phi)+R(-\Gamma\Phi+R\Phi)$$ $$+(-\Gamma R-R\Gamma+R^2)\Phi
=\Gamma^2\Phi+2R_x\Phi=\Gamma^2\Phi+i\Phi\Phi^T\La\Phi.$$
On the other hand (\ref{q1}) yields
$$\Phi_{xx}=\La^2\Phi+i\Phi\Phi^T\La\Phi,$$
which implies $\Gamma=\La.$
Therefore we have
\begin{equation}
\label{q8}
\Phi_{x}=-\La \Phi+R\Phi, 
\end{equation}
\begin{equation}
\label{q9}
R_x=\frac i2\Phi\Phi^T\La=-\La R-R\La+R^2. 
\end{equation}

To solve (\ref{q8}), we first consider the linear system
$$\Psi_x=-\La\Psi. $$
It is easy to see that
$$\Psi=(\alpha_1(t)e^{-\la_1x},..., \alpha_N(t)e^{-\la_Nx})^T.$$
Take  the solution of (\ref{q8}) to be of the form
\begin{equation}
\label{q12}
\Phi=(I-M)\Psi, 
\end{equation}
then $M$ has to satisfy
\begin{equation}
\label{q10}
M_x=M\La-\La M-R+RM. 
\end{equation}
 Comparing (\ref{q10}) with (\ref{q9}), one finds
\begin{equation}
\label{q11}
M=\frac 12 R\La^{-1}=\frac i4\p_x^{-1}(\Phi\Phi^T). \end{equation}
Equation (\ref{q12}) implies that
\begin{equation}
\label{q12a}
\Psi=\sum_{n=0}^{\infty}M^n\Phi. 
\end{equation}
By using (\ref{q11}) and (\ref{q12a}), it is found from that
$$\frac i4\p_x^{-1}(\Psi\Psi^T)
=\frac i4\p_x^{-1}\sum_{n=0}^{\infty}\sum_{l=0}^{n}M^l\Phi\Phi^T M^{n-l}$$
$$=\p_x^{-1}\sum_{n=0}^{\infty}\sum_{l=0}^{n}M^lM_xM^{n-l}
=\sum_{n=1}^{\infty}M^n.$$
Set
$$V=(V_{kj})=\frac i4\p_x^{-1}(\Psi\Psi^T), \qquad
V_{kj}=- \frac i4\frac {\alpha_k(t)\alpha_j(t)}{\la_k+\la_j}e^{-(\la_k+ \la_j)x},$$
one obtain
\begin{equation}
\label{q13}
(I+V)\Phi=\Psi,\qquad\text {or}\qquad \Phi=(I-M)\Psi=(I+V)^{-1}\Psi. \end{equation}

Notice that (\ref{q1}) and (\ref{q8}) gives rise to
\begin{equation}
\label{q14}
\La\Phi^*=(\La-R-\frac i2q)\Phi.
\end{equation}
 By inserting (\ref{q9}) and (\ref{q14}), (\ref{q2}) reduces to
$$\Phi_t=[-\La^{2}(\La-R-\frac i2q)-\frac i2q\La^2+(\La-R-\frac i2q)(-\La R-R\La+R^2)$$
\begin{equation}
\label{q15}
-(-\La R-R\La+R^2)(\La-R-\frac i2q)]
\Phi=-\La^3\Phi+R\La^{2}\Phi.
\end{equation}

Let $\Psi$ satisfy the linear system
\begin{equation}
\label{q16}
\Psi_t=-\La^3\Psi, 
\end{equation}
then
\begin{equation}
\label{q17}
\Psi=(\alpha_1(t)e^{-\la_1x},..., \alpha_N(t)e^{-\la_Nx})^T,
\qquad \alpha_i(t)=\beta_je^{-\la_j^3t}, \quad j=1,...,N.
\end{equation}

We now show that $\Phi$ determined by (\ref{q13}) and (\ref{q17}) satisfy (\ref{q15}). 
In fact, we have
$$\Phi_t=-(I+V)^{-1}\frac i4\p_x^{-1}(\Psi_t\Psi^T+\Psi\Psi_t^T)(I+V)^{-1}\Psi
+(I+V)^{-1}\Psi_t$$
$$=(1-M)(\La^3V+V\La^3)\Phi-(1-M)\La^3(1+V)\Phi=-\La^3\Phi+(I-M)V\La^{3}\Phi+M\La^3\Phi $$
$$=-\La^3\Phi+2M\La^3\Phi=-\La^3\Phi+R\La^2\Phi. $$
Therefore $\Phi$ given by (\ref{q13}) and (\ref{q17}) satisfy (\ref{q1}) and (\ref{q2}), 
simultaneously, 
and $q=\Phi^T\Phi^*$ is the solution of mKdV equation (\ref{s6}).  Notice that
$$\p_x (\Psi^T\Phi)= -\Psi^T\La\Phi+\Psi^T(-\La+R)\Phi$$
$$=\Psi^T(-2I+2M)\La\Phi=-2\Phi^T\La\Phi,$$
$$q_x=\frac 12(\Phi_x^T\Phi^*+\Phi^T\Phi^*_x)=\frac 12[(-{\Phi^*}^T\La-\frac i2q\Phi^T)\Phi^*
+\Phi^T(-\La\Phi+\frac i2q\Phi^*)]$$
$$=-\frac 12 ({\Phi^*}^T\La\Phi^*+\Phi^T\La\Phi)=-\text{Re}(\Phi^T\La\Phi).$$
So we have 
\begin{equation}
\label{q18}
q=\frac 12\text{Re}(\Psi^T\Phi)=\frac 12 \text{Re}\sum_{k=1}^N \alpha_k(t)e^{-\la_kx}\phi_k.
\end{equation}
Finally, as pointed out in \cite{1}, formulas (\ref{q13}) and (\ref{q18}) gives rise to  
the well-known
 $N$-soliton solution of mKdV equation (\ref{s6})
$$u=2\p_x\text{Im}\text{ln}(\text {det}(I+V)).$$

\hskip\parindent

\section{Conclusion}
\setcounter{equation}{0}
\hskip\parindent
We first factorize the mKdV equation into two commuting integrable $x$- and $t$-constrained
flows, then use them and their Lax representation to directly derive  the $N$-soliton solution 
for mKdV equation. The method proposed in present paper and previous paper \cite{19} 
provides a general procedure to construct $N$-soliton solution for soliton equations via their
$x$- and $t$-constrained flows and can be applied to other soliton eqqquations. 

\hskip\parindent

\section*{ Acknowledgment }
This work was in part supported by a grant from  City University of Hongkong 
(Project No. 7001072)  and  by 
the Special Funds for Chinese Major Basic Research Project "Nonlinear Science".

\hskip\parindent

\end{document}